\documentstyle[12pt,epsf]{article}

\newcommand{\gsim}{\;\rlap{\lower 3.5 pt \hbox{$\mathchar \sim$}} \raise 1pt
 \hbox {$>$}\;}
\newcommand{\lsim}{\;\rlap{\lower 3.5 pt \hbox{$\mathchar \sim$}} \raise 1pt
 \hbox {$<$}\;}

\renewcommand{\thefootnote}{\fnsymbol{footnote}}

\begin{document}    

\title{\vskip-3cm{\baselineskip14pt
\centerline{\normalsize\hfill MPI/PhT/97--032}
\centerline{\normalsize\hfill hep-ph/9706462}
\centerline{\normalsize\hfill May 1997}
}
\vskip1.5cm
Complete QCD Corrections of Order ${\cal O}(\alpha_s^3)$
to the Hadronic Higgs Decay}
\author{
 K.G.~Chetyrkin$^{a,b}$
 and 
 M.~Steinhauser$^{a}$
}
\date{}
\maketitle

\begin{center}
$^a${\it Max-Planck-Institut f\"ur Physik,
    Werner-Heisenberg-Institut,\\ D-80805 Munich, Germany\\ }
  \vspace{3mm}
$^b${\it Institute for Nuclear Research, Russian Academy of Sciences,\\
   Moscow 117312, Russia\\}
\end{center}

\begin{abstract}
\noindent 
We consider the decay of an intermediate
mass Higgs boson into hadrons up to order ${\cal O}(\alpha_s^3)$.  The
results from the diagrams containing only light degrees of freedom
were recently computed in a previous work by K.G. Ch.  
(Phys. Lett.  B 390 (1997) 309).
In this letter the remaining contributions
involving the top quark are evaluated analytically using an effective
field theory approach.  Coupled with the previous result the present 
calculation determines the complete next-next-to-leading 
order correction to the hadronic decay
rate of an intermediate mass Higgs  boson.
\end{abstract}

\vspace{3em}


\renewcommand{\thefootnote}{\arabic{footnote}}
\setcounter{footnote}{0}


The Higgs boson is the only still missing particle in the
standard model of particle physics. Up to now
only mass limits exist from the failure of experiments
at the CERN Large Electron-Positron Collider (LEP1)
to observe the process $e^+e^-\to ZH$. Currently 
the mass range $M_H\le65.6$~GeV is ruled out at the 95\%
confidence level \cite{Jan96}.

From the theoretical side a lot of effort has been invested to get
more insight into the properties of the Higgs boson
(for a review see \cite{Kni94}).  
Of particular
interest is thereby the intermediate mass range where $m_b \ll
M_H\lsim 2M_W$.  Then the dominant decay mode is the one into bottom
quarks.  Concerning QCD corrections the full mass dependence at ${\cal
O}(\alpha_s)$ was evaluated in \cite{DreHik90}.  However, especially
for the fermionic decay of an intermediate mass Higgs boson it turns
out that an expansion in the quark masses provides reasonable
approximations.  At ${\cal O}(\alpha_s^2)$ besides the massless limit
(i.e. keeping only the overall factor $m_b^2$) \cite{GorKatLarSur90}
also subleading mass corrections in the $m_b^2/M_H^2$ expansion are
known \cite{Sur94,CheKwi96}.

Recently the contribution to the scalar  polarization function
of the Higgs boson containing only light quarks was considered
at four loops \cite{Che97}. Its imaginary part was evaluated
analytically in the massless limit leading to corrections of 
${\cal O}(\alpha_s^3)$ to the Higgs decay rate. The corresponding 
corrections prove to be numerically more important than the power suppressed
contribution of ${\cal O}(\alpha^2_s m_b^2/M_H^2)$.

In Ref.~\cite{CheKwi96} additional quasi-massless (and numerically
important) contributions of order $\alpha_s^2$ have been identified
and elaborated.  They come from so-called singlet diagrams with a
non-decoupling top quark loop 
inside\footnote{Similar effects for the decay of
  the Z-boson to quarks were first discovered in \cite{KniKue90b}.}. 
The results of Ref.~\cite{CheKwi96} have been confirmed and extended by
the computation of power-suppressed terms of order
$\alpha_s^2(M_H^2/M_t^2)^n\, (n=1,2,\ldots)$ \cite{LarRitVer95}.
For completeness we should mention that
the exact result for the imaginary part of the double-bubble diagram
with massless external quark and internal top quark can be found in
\cite{Kni95}.  
Radiative corrections enhanced by a factor $M_t^2$,
usually expressed through $X_t=G_FM_t^2/8\pi^2\sqrt{2}$,
are also available up to the three-loop order
\cite{KniSte95,CheKniSte97} and will be compared at the end with the 
new terms of ${\cal O}(\alpha_s^3)$.

In this letter we calculate the top-induced corrections at ${\cal
O}(\alpha_s^3)$ and hence complete the analysis of the hadronic Higgs
decay at NNNLO, as far as one neglects power suppressed corrections.
Although the phenomenological interesting decay is the one into bottom
quarks the following discussion will be kept more general and a
generic light quark $q$ will be considered.  However, for the
numerical discussion we will come back to the case of bottom quarks.

We start with 
the bare Yukawa Lagrangian,
\begin{equation}
{\cal L}_Y = -\frac{H^0}{v^0}
\left(
 \sum_q m_q^0 \bar{q}^0 q^0 + m_t^0 \bar{t}^0 t^0
\right)
{},
\end{equation}
where $v$ is the Higgs vacuum-expectation value and
the superscript 0 labels bare quantities.
Assuming that the Higgs boson  mass, $M_H$ is less than the
top quark mass $m_t$, ${\cal L}_Y$  can be replaced by an 
effective Lagrangian  produced by integrating out  the top
quark field.  According to Refs.~\cite{InaKubOka83,CheKniSte97,CheKniSte972}
the resulting Lagrangian reads
\begin{eqnarray}
{\cal L}_Y^{\rm eff} &=& -\frac{H^0}{v^0}\left[
C_1 \left[O_1^\prime\right]
+\sum_q
C_{2q} \left[O_{2q}^\prime\right]
\right],
\label{eqlageff}
\end{eqnarray}
where $[O_1^\prime]$ and $[O_{2q}^\prime]$ are the 
renormalized counterparts of 
the bare operators
\begin{eqnarray}
O_1^\prime\,\,=\,\,\left(G_{a\mu\nu}^{0\prime}\right)^2,
&&
O_{2q}^\prime\,\,=\,\,m_q^{0\prime}\bar q^{0\prime}q^{0\prime},
\nonumber
\end{eqnarray}
with $G_{a\mu\nu}^{0\prime}$ being the (bare) field strength tensor of
the gluon.  The primes mark the quantities defined in the effective
$n_f = 5 $ QCD including only light (in comparison to the top quark) 
$u,d,s,c$
and $b$ quarks. All the dependence on the top quark gets localized in
the coefficient functions $C_1$ and $C_{2q}$.  If the latter are
known  the computation of the total decay rate 
$\Gamma_H = \Gamma(H \to\mbox{hadrons})$ is reduced
to the evaluation of the functions 
$\Delta_{jk}(M_H^2)$ ($jk = 11, 12, 22 $)
related via the optical theorem to the absorptive parts 
of the scalar correlators 
\begin{equation}
\label{deltaij}
\Pi_{jk}(q^2) = 
i\int dx e^{iqx}\langle 0|\;T[\; O^\prime_j(x)
O^\prime_k(0)\,]\;|0\rangle\bigg|_{q^2=M_H^2}
{}.
\end{equation}
It should be stressed that by the very meaning of the effective
Lagrangian (\ref{eqlageff}) the correlators of Eq.~(\ref{deltaij}) 
may be computed within the effective massless QCD, 
which leads to a drastic
simplification of the calculation. In addition, the coefficient
functions $C_1$ and $C_{2q}$ essentially depend on only one kinematical
variable, the top quark mass, which also simplifies their calculation a lot. 

\begin{figure}[t]
 \begin{center}
   \leavevmode
   \epsfxsize=14.0cm
   \epsffile[125 640 500 730]{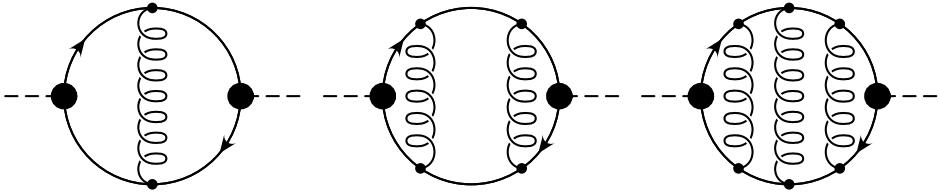}
 \end{center}
\caption{\label{figdel22}
         Typical Feynman diagrams contributing to $\Delta_{22}^q$. 
         The solid circles represent the operator
         $\left[O_{2q}^\prime\right]$.}
\end{figure}

\begin{figure}[t]
 \begin{center}
   \leavevmode
   \epsfxsize=14.0cm
   \epsffile[130 640 500 730]{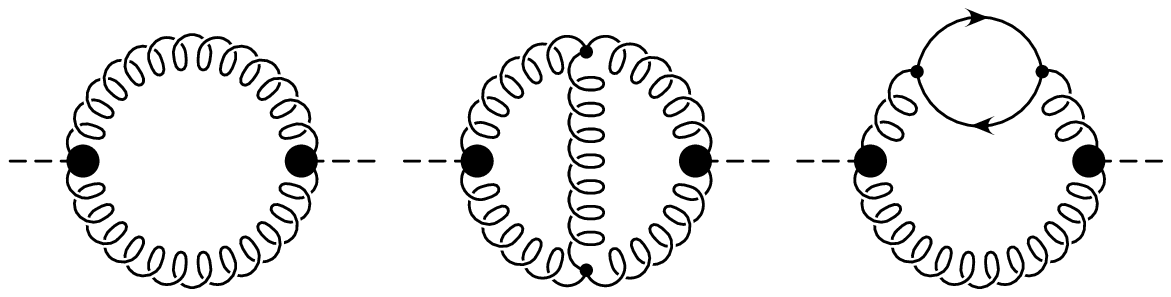}
 \end{center}
\caption{\label{figdel11}
         Typical Feynman diagrams contributing to $\Delta_{11}^q$. 
         The solid circles represent the operator
         $\left[O_1^\prime\right]$.}
\end{figure}

\begin{figure}[t]
 \begin{center}
   \leavevmode
   \epsfxsize=15.0cm
   \epsffile[70 630 570 720]{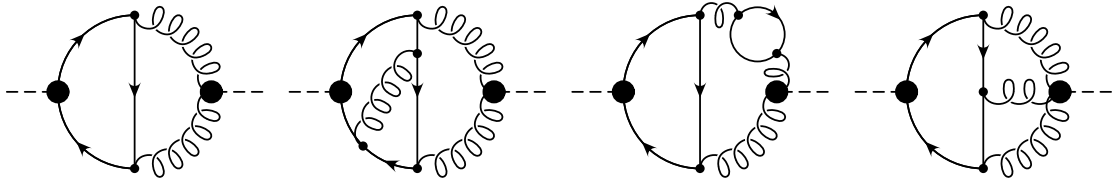}
 \end{center}
\caption{\label{figdel12}. Two- and some of the 
                         three-loop diagrams contributing
                         to $\Delta_{12}^q$. 
                         The solid circles represent the operators
                         $\left[O_1^\prime\right]$ and 
                         $\left[O_{2q}^\prime\right]$, respectively.
                         }
\end{figure}

Typical diagrams contributing to $\Delta_{22}^q$, $\Delta_{11}$ and
$\Delta_{12}^q$ are shown in Figs.~\ref{figdel22}-\ref{figdel12}. 
Note that every massless
diagram contributing to $\Delta_{22}^q$ to order $\alpha_s^3$ does only 
have cuts containing at least one quark-antiquark pair.  This
allows one to unambiguously associate $\Delta^q_{22}$ to 
$\Gamma(H \to q \bar{q})$  --- the  partial decay rate
of the Higgs boson to  hadrons containing at least one $q\bar{q}$ pair. 
The diagrams contributing to $\Delta_{11}$ (see Fig.~\ref{figdel11})
describe the production of gluons.
The leading order diagram has clearly only contributions to the 
$gg$ final state. Starting from NLO, however, there are contributions
from diagrams leading both to $gg$, $ggg$ and $gq\bar{q}$ final states. 
The interpretation of the  $\Delta_{12}^q$ (see Fig.~\ref{figdel12})
is even  more complicated. Here different final states begin 
to appear already in  the leading non-vanishing order
\cite{CheKwi96}.
In this paper we will adopt a pragmatical point of view 
and evaluate the total hadronic decay rate without any attempt to
differentiate between final states
\cite{DjoSpiZer96}. 

The functions $\Delta_{11}$, $C_1$ and $\Delta_{22}^q$ have been
computed analytically in \cite{InaKubOka83,Spira,CheKniSte972} 
and \cite{Che97}. Thus, to complete the  evaluation of 
$\Gamma_H$ in the ${\cal O}(\alpha_s^3)$ approximation one
needs to compute $C_{2q}$ as well as 
$\Delta_{12}^q$. This calculation and its
results are described below.

For convenience we
should recall the relevant formulae which connect the bare and
renormalized quantities appearing in Eq.~(\ref{eqlageff}).
For the operators we have
\begin{eqnarray}
\left[O_1^\prime\right]&=&
\left[1+2\left(\frac{\alpha_s\partial}{\partial\alpha_s}\ln Z_g\right)\right]
O_1^\prime
-4\left(\frac{\alpha_s\partial}{\partial\alpha_s}\ln Z_m\right)
\sum_q O_{2q}^\prime,
\nonumber\\
\left[O_{2q}^\prime\right]&=&O_{2q}^\prime,
\end{eqnarray}
and for the coefficient function the relations look like:
\begin{eqnarray}
C_1&=&\frac{1}{1+2(\alpha_s\partial/\partial\alpha_s)\ln Z_g}C_1^0,
\nonumber\\
C_{2q}&=&\frac{4(\alpha_s\partial/\partial\alpha_s)\ln Z_m}
{1+2(\alpha_s\partial/\partial\alpha_s)\ln Z_g}C_1^0+C_{2q}^0,
\end{eqnarray}
where the renormalization constant $Z_g$ and $Z_m$ are defined in
the effective theory. In the order we are interested in these constants
are needed up the one- and two-loop level, respectively,
because $C_1^0$ is already proportional to $\alpha_s$.

\begin{figure}[t]
 \begin{center}
   \leavevmode
   \epsfxsize=15.0cm
   \epsffile[70 620 560 720]{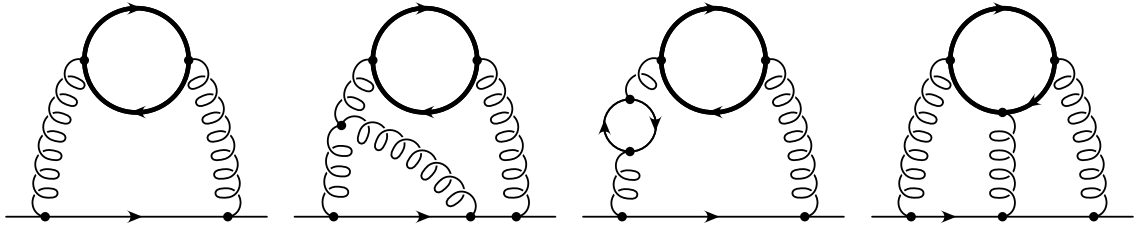}
 \end{center}
\caption{\label{figsig} Two- and some of the 
                        three-loop diagrams contributing to
                        $\Sigma_S^{0t}$ and $\Sigma_V^{0t}$. 
                        The thick (thin) lines 
                        represents the top quark (light quarks).
                        }
\end{figure}

For our purpose we need $C_1$ up to order $\alpha_s^2$
which may be found in
\cite{InaKubOka83,Spira,CheKniSte972}.
$C_{2q}$ is also known up to ${\cal O}(\alpha_s^2)$
\cite{CheKniSte97}
but for the present analysis needed up to ${\cal O}(\alpha_s^3)$.
To this aim we want to use the low energy theorem (LET) 
which allows to attach the Higgs boson via differentiation
w.r.t. the masses involved in the process. The formula given in
\cite{CheKniSte97} for the computation of $C_{2q}^0$ simplifies in our
case to
\begin{eqnarray}
C_{2q}^0 &=& 1 - \frac{m_t^0\partial}{\partial m_t^0}
\left(\Sigma_S^{0t}(0)+\Sigma_V^{0t}(0)\right).
\end{eqnarray}
It is understood that after the derivative w.r.t. the bare top mass
is done the renormalization of the parameters $m_t$ and $\alpha_s$ 
is performed.
The superscript $t$ indicates that only the diagrams where
at least one top quark is present have to be taken into account.
The first non-vanishing contribution arises at two-loop level
involving one diagram.
At three loops altogether 25 diagrams contribute.
Some typical examples are depicted in Fig.~\ref{figsig}.
Note that only the pole parts have to be computed because the
integrals are logarithmically divergent and consequently
proportional to $(\mu^2/m_t^2)^\varepsilon$. The differentiation
w.r.t. the top mass then leads to an additional factor $\varepsilon$.
The diagrams are generated with the program QGRAF
\cite{Nog93} and fed into the package MATAD
written in FORM
\cite{FORM} for the purpose to solve three-loop tadpole integrals.
The calculation was performed with arbitrary QCD gauge parameter $\xi$.
$\Sigma_S^{0t}$ and $\Sigma_V^{0t}$ separately still depend on $\xi$,
however, the final result for $C_{2q}$ is 
independent of the gauge parameter which
serves as a welcome check.
Expressing the result in terms of the $\overline{\mbox{MS}}$ 
top mass, $m_t$, we get
\begin{eqnarray}
C_{2q} &=&
1
+\left(\frac{\alpha_s^{(6)}(\mu)}{\pi}\right)^2\left[
\frac{5}{18}
-\frac{1}{3}\ln\frac{\mu^2}{m_t^2}
\right]
+\left(\frac{\alpha_s^{(6)}(\mu)}{\pi}\right)^3\left[
\frac{311}{1296} 
+ \frac{5}{3}\zeta(3)
\right.
\nonumber
\\&&
\left.\quad\mbox{}
- \frac{175}{108}\ln\frac{\mu^2}{m_t^2}
- \frac{29}{36}\ln^2\frac{\mu^2}{m_t^2}
+ n_l\left(
  \frac{53}{216}
 +\frac{1}{18}\ln^2\frac{\mu^2}{m_t^2}
\right)
\right],
\end{eqnarray}
with $\zeta(3)\approx1.20206$.
$n_l$ is the number of light quarks.
Using the relation between the $\overline{\mbox{MS}}$
and the on-shell mass, $M_t$, at one-loop level,
$m_t/M_t = 1-(4/3+\ln\mu^2/M_t^2)\alpha_s/\pi$,
one gets:
\begin{eqnarray}
C_{2q}^{OS} &=&
1
+\left(\frac{\alpha_s^{(6)}(\mu)}{\pi}\right)^2\left[
\frac{5}{18}
-\frac{1}{3}\ln\frac{\mu^2}{M_t^2}
\right]
+\left(\frac{\alpha_s^{(6)}(\mu)}{\pi}\right)^3\left[
-\frac{841}{1296} 
+ \frac{5}{3}\zeta(3)
\right.
\nonumber
\\&&
\left.\quad\mbox{}
- \frac{247}{108}\ln\frac{\mu^2}{M_t^2}
- \frac{29}{36}\ln^2\frac{\mu^2}{M_t^2}
+ n_l\left(
  \frac{53}{216}
 +\frac{1}{18}\ln^2\frac{\mu^2}{M_t^2}
\right)
\right],
\end{eqnarray}
For completeness we also list the result for $C_1$
\cite{InaKubOka83,Spira,CheKniSte972}:
\begin{eqnarray}
C_1 &=& 
-\frac{1}{12}\,\frac{\alpha_s^{(6)}(\mu)}{\pi}
\left[
1+\frac{\alpha_s^{(6)}(\mu)}{\pi}
\left(\frac{11}{4}-\frac{1}{6}\ln\frac{\mu^2}{m_t^2}
\right)
\right].
\end{eqnarray}

On the basis of the effective Lagrangian of Eq.~(\ref{eqlageff})
it is possible to write down the decay rate of the Higgs boson into
hadrons in the following form:
\begin{eqnarray}
\label{eqgammahqq}
\Gamma\left(H\to \mbox{hadrons}\right)
&=&
\left(1+\delta_u\right)^2
\bigg\{
\sum_q
A_{q\bar q}
\left[
\left(1+\Delta^q_{22} \right)\left(C_{2q}\right)^2
+\Delta^q_{12}\,C_1C_{2q}
+
\Delta^{\rm hdo}
\right]
\nonumber\\&&\mbox{}\qquad\qquad
+
A_{gg}
\,\Delta_{11}\,\left(C_1\right)^2
\bigg\},
\label{eqgamhad}
\end{eqnarray}
with
$A_{q\bar q}=3G_FM_Hm_q^2/4\pi\sqrt2$
and 
$A_{gg}=4G_FM_H^3/\pi\sqrt{2}$.
The universal corrections $\delta_u$ arise from the renormalization
of the factor $H^0/v^0$. The terms up to ${\cal O}(\alpha_s^2X_t)$
can be found in \cite{KniSte95}.
$\Delta^{\rm hdo}$ summarizes the corrections coming from
{\bf h}igher {\bf d}imensional {\bf o}perators. 
They are at least suppressed by a factor
$\alpha_s^2M_H^2/M_t^2$. 
The factors $\Delta_{ij}$ contain the electroweak and QCD
corrections from the light degrees of freedom only. 
As we are interested up to an accuracy of ${\cal O}(\alpha_s^3)$ 
$\Delta_{11}$ and $\Delta^q_{22}$ are known.
However, $\Delta_{12}^q$ is only available at the two-loop level where one
diagram has to be evaluated (see Fig.~\ref{figdel12}). 
We have to extend the analysis 
to three loops where altogether 28 massless three-loop
diagrams have to be computed. 
This was done with the help of the
program package MINCER
\cite{MINCER}.
In Fig.~\ref{figdel12} some graphs are pictured.
The single diagrams again depend on $\xi$ which cancels in the proper sum.
The result reads:
\begin{eqnarray}
\Delta^q_{12} &=& 
\frac{\alpha_s^{(5)}(\mu)}{\pi}
\left[
-\frac{92}{3}
-8\ln\frac{\mu^2}{M_H^2}
\right]
\nonumber\\
&&
+\left(\frac{\alpha_s^{(5)}(\mu)}{\pi}\right)^2
\left[
- \frac{15073}{18} 
+ 76\zeta(2)
+ 156\zeta(3)
- \frac{1028}{3}\ln\frac{\mu^2}{M_H^2}
- 38\ln^2\frac{\mu^2}{M_H^2} 
\right.
\nonumber\\
&&
\left.\mbox{}
+ n_l\left(
  \frac{283}{9}
- \frac{8}{3}\zeta(2)
- \frac{16}{3}\zeta(3)
+ \frac{112}{9}\ln\frac{\mu^2}{M_H^2}
+ \frac{4}{3}\ln^2\frac{\mu^2}{M_H^2}
\right)
\right],
\end{eqnarray}
with $\zeta(2)=\pi^2/6$.
The ${\cal O}(\alpha_s)$ terms can also be found in
\cite{CheKwi96,LarRitVer95,CheKniSte97}, 
the ${\cal O}(\alpha_s^2)$ terms are new.

Let us now compare the new results of ${\cal O}(\alpha_s^3)$
with the previous ones and also with other known correction
terms which might be of the same order of magnitude.
Here, we have in mind terms of order
$\alpha_s^2m_b^2/M_H^2$,
$\alpha\alpha_s$,
$\alpha_s^2X_t$
and
$\alpha_s^2M_H^2/M_t^2$.
Thereby it is convenient to write Eq.~(\ref{eqgamhad}) in the 
form\footnote{
  From now we will consider all quarks with mass lighter than
  $m_b$ as massless. This means that the sum in
  Eq.~(\ref{eqgamhad}) reduces to $q=b$. Furthermore the
  numerical discussion comparing the new corrections with previously
  known terms is restricted to the part proportional to 
  to $A_{b\bar{b}}$.}:
\begin{eqnarray}
\Gamma(H\to\mbox{hadrons}) &=& 
A_{b\bar{b}}\left(
1+\Delta^b_l+\Delta_t
\right)
+
\left(1+\delta_u\right)^2
A_{gg}
\,\Delta_{11}\,\left(C_1\right)^2,
\end{eqnarray}
where $\Delta_l^b$ contains only corrections form light 
degrees of freedom and all top-induced terms from the 
first line of Eq.~(\ref{eqgamhad}) are contained in $\Delta_t$.
Furthermore we express $\Delta_t$ also in terms of $\alpha_s^{(5)}(\mu)$.
Choosing $\mu^2=M_H^2$ and $n_l=5$ we find 
\begin{eqnarray}
\Delta_l^b &=&
- 6 \frac{(m_b^{(5)})^2}{M_H^2}
+ 0.472 \,\frac{\bar\alpha(M_H)}{\pi}
+ 0.651 \,\frac{\bar\alpha(M_H)}{\pi}a_H^{(5)}
+ a_H^{(5)}\left(
   5.667 
   - 40.000\frac{(m_b^{(5)})^2}{M_H^2}
\right)
\nonumber\\&&\mbox{}
+\left(a_H^{(5)}\right)^2\left( 
29.147 
- 87.725 \frac{(m_b^{(5)})^2}{M_H^2}
\right)
+ 41.758 \left(a_H^{(5)}\right)^3,
\label{eqdell}
\\
\Delta_t &=&
\left(a_H^{(5)}\right)^2
\left(
3.111
-0.667\,L_t
+\frac{(m_b^{(5)})^2}{M_H^2}
\left(
-10
+4\,L_t
+\frac{4}{3}\ln\frac{(m_b^{(5)})^2}{M_H^2}
\right)
\right) 
\nonumber\\&&\mbox{}
+\left(a_H^{(5)}\right)^3
\left(
50.474
-8.167\,L_t
-1.278\,L_t^2
\right)
+ \left(a_H^{(5)}\right)^2\frac{M_H^2}{M_t^2}
\left(
0.241 
- 0.070\, L_t
\right)
\nonumber\\&&\mbox{}
+X_t\left(1
- 4.913 a_H^{(5)}
+ \left(a_H^{(5)}\right)^2
\left(
-72.117
-20.945\,L_t
\right)
\right),
\label{eqdelt}
\end{eqnarray}
with $a_H^{(5)}=\alpha_s^{(5)}(M_H)/\pi$ and  $L_t=\ln M_H^2/M_t^2$.
The term in Eq.~(\ref{eqdelt}) proportional to $M_H^2/M_t^2$
is the leading contribution 
from $\Delta^{\rm hdo}$. 
The $(m_b^{(5)})^2/M_H^2$ corrections in $\Delta_t$ arise from
the singlet diagram with one top and one bottom quark triangle
and can be found in \cite{CheKwi96}.
At this point we should mention that still 
contributions with pure gluonic final states 
are contained in Eq.~(\ref{eqgamhad}).
At ${\cal O}(\alpha_s^3)$, however, these
corrections are not yet known and can thus not be subtracted.

In the approximation considered in this paper we have
$-2\lsim L_t<0$. This means that the logarithm needs not 
necessarily to be resummed as in addition the coefficients
in front of $L_t$ are much smaller than the constant term.

One observes that the new top-induced corrections of
${\cal O}(\alpha_s^3)$ are numerically of the same size as 
the previous ones arising from ``pure'' QCD.
Furthermore one should mention that the coefficient of the
$M_t$-suppressed terms are tiny and, as
$\alpha_s/X_t\approx30$, also the $\alpha_s^2X_t$ enhanced terms 
are less important than the cubic QCD corrections.
For comparison in Eq.~(\ref{eqdell}) also the two-loop
corrections of order $\alpha\alpha_s$ are listed.
In principle also higher order mass corrections are available 
\cite{HarSte97}. However, in the case of bottom quarks
it turns out that they are tiny.

To summarize, in this letter the top-induced corrections of order
$\alpha_s^3$ for the hadronic Higgs boson decay were presented. An
effective Lagrangian was constructed and both the coefficient
functions and relevant correlators were evaluated up to three
loops. The new results combined with the ones in \cite{Che97} lead to
complete corrections of ${\cal O}(\alpha_s^3)$ to the hadronic Higgs
decay.


\centerline{\bf Acknowledgments}
\smallskip\noindent
We would like to thank B.A. Kniehl for useful discussions.
This work was supported by INTAS under Contract INTAS-93-744-ext.



\begin{thebibliography}{99}

\bibitem{Jan96} 
P. Janot,
in {\it Proceedings of the Ringberg Workshop: The Higgs Puzzle---What can we
learn from LEP2, LHC, NLC, and FMC?}, Rottach-Egern, Germany, 8--13 December 
1996, edited by B.A. Kniehl (World Scientific, Singapore, in print).

\bibitem{Kni94}
B.A. Kniehl, {\it Phys. Rept.} {\bf 240} (1994) 211.

\bibitem{DreHik90}
M. Drees and K. Hikasa, {\it Phys. Lett.} {\bf B 240} (1990) 455;
{\bf B 262} (1991) 497 (E).

\bibitem{GorKatLarSur90} 
S.G. Gorishny, A.L. Kataev, S.A. Larin, and L.R. Surguladze,
{\it Mod.\ Phys.\ Lett.} {\bf A 5} (1990) 2703; 
{\it Phys.\ Rev.} {\bf D 43} (1991) 1633.

\bibitem{Sur94}
L.R. Surguladze, {\it Phys.\ Lett.} {\bf B 341} (1994) 60.

\bibitem{CheKwi96}
K.G. Chetyrkin and A. Kwiatkowski, {\it Nucl. Phys.} {\bf B 461} (1996) 3.

\bibitem{Che97}
K.G. Chetyrkin, {\it Phys. Lett.} {\bf B 390} (1997) 309.

\bibitem{KniKue90b}
B.A. Kniehl and J.H. K\"uhn,
{\it Phys. Lett.} {\bf B 224} (1989) 229;
{\it Nucl. Phys.} {\bf B 329} (1990) 547.

\bibitem{LarRitVer95}
S.A. Larin, T. van Ritbergen and J.A.M. Vermaseren,
{\it Phys. Lett.} {\bf B 362} (1995) 134.

\bibitem{Kni95}
B.A. Kniehl, {\it Phys. Lett.} {\bf B 343} (1995) 299.

\bibitem{KniSte95} 
B.A. Kniehl and M. Steinhauser,
{\it Nucl.\ Phys.} {\bf B 454} (1995) 485; 
{\it Phys.\ Lett.} {\bf B 365} (1996) 297.

\bibitem{CheKniSte97}
K.G. Chetyrkin, B.A. Kniehl and M. Steinhauser,
{\it Phys. Rev. Lett.} {\bf 78} (1997) 594;
{\it Nucl. Phys.} {\bf B 490} (1997) 19.

\bibitem{InaKubOka83}
T. Inami, T. Kubota, and Y. Okada, {\it Z. Phys.} {\bf C 18} (1983) 69.

\bibitem{CheKniSte972}
K.G. Chetyrkin, B.A. Kniehl and M. Steinhauser, MPI/PhT/97-006,
hep-ph/9705240 (to be published in {\it Phys. Rev. Lett.}).

\bibitem{DjoSpiZer96}
A. Djouadi, M. Spira, and P.M. Zerwas, {\it Z. Phys.} {\bf C 70} (1996) 427.

\bibitem{Spira}
A. Djouadi, M. Spira, and P.M. Zerwas, 
{\it Phys.\ Lett.} {\bf B 264} (1991) 440.

\bibitem{Nog93} 
P. Nogueira,
{\it J. Comput.\ Phys.} {\bf 105} (1993) 279. 

\bibitem{FORM}
J.A.M. Vermaseren, {\it Symbolic Manipulation with FORM},
(Computer Algebra Netherlands, Amsterdam, 1991).

\bibitem{MINCER} 
S.A. Larin, F.V. Tkachev, and J.A.M. Vermaseren,
NIKHEF Report No.\ NIKHEF--H/91--18 (September 1991).

\bibitem{HarSte97}
R. Harlander and M. Steinhauser,
MPI/PhT/97-013, TTP-97-12, hep-ph/9704436.

\end{thebibliography}
\end{document}